# Atomistic modeling of the anomalous helium behaviors in vanadium

Nengwen Hu,[1,2] Canglong Wang,[1,*] Huiqiu Deng,[3] Shifang Xiao,[3] Chengbin Wang,[4] Lei Yang[1] and Wangyu Hu[2,*]

[1]Institute of Modern Physics, Chinese Academy of Sciences, Lanzhou 730000, China

[2]College of Materials Science and Engineering, Hunan University, Changsha 410082, China

[3]School of Physics and Electronics, Hunan University, Changsha 410082, China

[4]Institute of Applied Physics, Chinese Academy of Sciences, Shanghai 200000, China

**Abstract**

*Ab initio* calculations have been performed to clarify the primary behaviors of He atoms in vanadium and to generate the database for the development of the interatomic potential for V-He system within the framework of the "s-band" model. The calculated formation energies of the tetrahedral, octahedral and substitutional He defects, as well as those for $He_2$, $He_3$ and $He_2V$ clusters are reasonable when compared with relevant experimental results and *ab initio* calculations under the same conditions. The applicability of the present V-He potential for atomistic simulations to investigate the moving, clustering, and trapping of interstitial He atoms in vanadium are demonstrated. Similar to the results in iron and tungsten, the interstitial He atoms can migrate quickly in vanadium. However, He atoms aggregate into clusters with low binding energies, and the trapping of He atoms depend much on the pre-existing traps in vanadium.

Corresponding author: wyhu@hnu.edu.cn; clwang@impcas.ac.cn

# 1. Introduction

There is an urgent need to develop new high-performance materials with good tolerance to high operating temperature and stress, long-term reliability under operating conditions, excellent fabricability and weldability, and reasonable production cost, which is very important for various applications such as aircraft turbine components to pressure vessel materials. As one of the high-performance materials, vanadium-base alloys have aroused tremendous attentions for many years. Due to their outstanding performances, such as reduced-activation, irradiation resistance and high temperature strength [1], vanadium alloys have a wide range of possible technological and engineering application. Particularly, the type of alloy could become candidates for high temperature structure materials and protective coatings in nuclear reactors.

Under irradiation environments, apart from the presence of vacancies, self-interstitial atoms (SIAs), and dislocation loops, impurity He atoms can be also introduced into materials. Furthermore, the accumulation of He atoms may induce the formation of He bubbles, which will lead to void swelling and low-temperature intergranular embrittlement, surface roughening, blistering, and premature creep rupture at high-temperatures [2]. To date, despite extensive efforts at the properties of vanadium alloys [3-5], there is a great deal of dispute regarding the effect of He atoms on the properties of vanadium alloys [6]. On one hand, it has currently been recognized that the impurity He atoms may severely degrade the properties of vanadium alloys. For example, the implantation of He ion could severely reduce the ductility of vanadium alloys, which directly determines the upper operating temperature of vanadium based alloy [7]. On the other hand, however, Kurtz *et al.* have experimentally observed in vanadium alloys that the neutron irradiation with the implantation of He ion influences little on the helium embrittlement [8]. At the same time, the implanted He may even improve the property of creep in the presence of small He clusters. Due to the limit of the 14 MeV neutron sources and relevant techniques, no further investigations on the confusions of the helium effect in vanadium alloys have been reported [1]. A deeper knowledge of the helium effect is technologically crucial for the optimization of desirable vanadium alloys. Fortunately, atomic scale simulation techniques have been widely proved to be effective in the irradiation and helium effects of structural materials [9-19]. The credibility of results delivered by atomistic simulations largely depends on the robustness of interatomic potentials employed in the simulations, which should be able to reproduce correctly various fundamental physical properties of relevant materials systems. Most simulations employ the best known many-body potentials based on the Finnis-Sinclair (F-S) or the embedded-atom method (EAM) [20-23]. To the present authors' knowledge, however, only a two-body interatomic potential fitted by Chernov et al [24] is available for the V-He system. Unfortunately, the pair interaction between vanadium and He atoms becomes negative at the distances between 2.9 and 5.3 Å, which indicates an unphysical attractive interaction, rather than repulsive. Nevertheless, it has recently recognized that there is little attractive interaction between vanadium and He atoms due to the closed shell structure of electrons of He atoms [25, 26]. Under its description, the interstitial He atoms tend to migrate between equivalent octahedral sites passing through a tetrahedral site with the diffusion barrier of 0.22 eV, which is higher than that from *ab initio* prediction (0.06 eV). Besides, the binding energy was overestimated for He atoms, which is inconsistent with that from experimental and *ab initio* calculations [24]. As a consequence, in order to gain insight into the essence of helium effect, an accurate interatomic potential is urgently needed in V-He system.

In the absence of experimental data, the data obtained from *ab initio* calculations are widely approved and referred in the development of interatomic potentials [27-30]. The damage of He atoms to materials derives from their aggregation. To evaluate the damage, we have conducted *ab initio* calculations to clarify the primary behaviors of He atom in vanadium. The database for the development of potential was obtained with *ab initio* calculations. Based on the database, we have developed an accurate V-He interatomic potential within the framework of "s-band" model by matching the database from *ab initio* calculations. With the present potential, the dynamical mechanisms (moving, binding, and trapping) relevant to the aggregation of He atoms have been investigated with molecular dynamics method. For comparison, the self-trapping behaviors in iron were also investigated to gain insight into the anomalous behaviors of He atoms in vanadium. The present paper is organized as follows: In Sec. 2, the bases of V-He potential, the fitting technology and all modeling parameters; In Sec. 3, the discussions on the fitted and predicted results; In Sec. 4, we have summarized the results and the possible competence applications of the present V-He potential.

## 2. Methodology

The purpose of *ab initio* calculations conducted in the present work is twofold: (1) confirm the anomalous clustering behaviors of He atoms in vanadium, and (2) generate a database for construction interatomic potential for the V-He system.

### 2.1 Ab initio calculations

The electronic structure calculations reported herein were performed using the Vienna *ab initio* simulation package (VASP) [31, 32]. The Kohn-Sham equations were solved self-consistently using a plane-wave basis set with projector augmented wave (PAW) pseudopotentials [33]. The exchange and correlation functionals were taken in the form proposed by Perdew and Wang (PW91) within the generalised gradient approximation (GGA) [34].

In all of the calculations, the minimal cutoff energy providing energy convergence was set to 400 eV. The Brilliouin zone integration was performed using a 5×5×5 k mesh within the Monkhorst-Pack scheme for the geometry optimization of V-He system. The self-consistent loop terminates when the total energy was converged to $1.0×10^{-4}$ eV per atom, and the maximum forces on each unconstrained atom are smaller than 0.01 eV/Å. We have employed $3a_0×3a_0×3a_0$ supercell for electronic properties and defect calculations, where $a_0$ is the lattice constant of vanadium (2.977 Å).

The formation energies of $He_nV_m$ (n He atoms and m vacancies) cluster in vanadium were calculated according to the following formula,

$$E_f(He_nV_m) = E_{tot}(He_nV_m) - [nE_{He}^C + (N-m)E_V^C], \qquad (1)$$

where $E_{tot}(He_nV_m)$ is the total energy of the system containing a defect cluster of $He_nV_m$. $E_{He}^C$ and $E_V^C$ represent the cohesive energy of a He atom in fcc structural and a vanadium atom in bcc lattice, respectively. In the present work, the binding energies of an individual He atom to defect clusters were calculated as following,

$$E_b(He) = E_f(He) + E_f(He_nV_m) - E_f(He_{n+1}V_m), \qquad (2)$$

where $E_f(He)$ means the formation energy of a He atom at tetrahedral interstitial sites of bcc structural vanadium.

**2.2 Potential construction**

To enable atomistic simulations of the V-He system, we need a semi-empirical potential that examines the binding properties and relative stabilities of He-V and interstitial He clusters. To this end, we have developed a V-He interatomic potential within the framework of "s-band" model by matching the results obtained from *ab initio* calculations. The pair and many-body interactions for V-V are described by the Finnis-Sinclair (FS) type potential [35], and the He-He interaction is taken as the Hartree-Fock-dispersion pair potential [36], which describes He properties in vacuum. Similar to the many-body potential formalism, the potential for the V-He interaction consists of a pair potential and an embedding function. In this method, the total energy of the V-He system is defined by the following expression in an orthogonal Cartesian system:

$$U = \sum_{i=1}^{N-1} \sum_{j=i+1}^{N} \varphi(r_{ij}) + \sum_{i=1}^{N} F_s(\rho_s^i), \qquad (3)$$

where the first term represents a repulsive pair potential and the second term is the many-body interaction or hybridization energy that provides the contribution from the s-band electron density due to hybridization between the d-electron of V atoms and the s-electron of He atoms. The pair potential $\varphi(r_{ij})$ is parameterized in the following form:

$$\varphi(r_{ij}) = \sum_{k=1}^{6} a_k (r_k - r)^3 H(r_k - r), \qquad (4)$$

where $H(r_k-r)$ is the Heaviside function. The potential parameters are $a_k$ and $r_k$ ($k$=1-6). The cutoff distance is set at 3.6 Å. The hybridization energy $F_s(\rho_s^i)$ is given in the following form:

$$F_s(\rho_s) = b_1\sqrt{\rho_s} + b_2\rho_s^2 + b_3\rho_s^4, \tag{5}$$

where $\rho_s$ is the hybridization electron density, with the potential parameters $b_1$, $b_2$, and $b_3$. We have chosen the 1s-type and 4s-type Slater functions for He and vanadium [37], respectively:

$$\chi^{1s} = N_{1s}\exp(-\xi_{1s}r), \tag{6}$$

$$\chi^{4s} = N_{4s}r^3\exp(-\xi_{4s}r), \tag{7}$$

The hybridization electron density for V-He system can be defined by:

$$\rho_s = \sum N_s\, r^3\exp(-2\xi_s r), \tag{8}$$

where $\xi_s$ is an average $\xi$ from the 1s and 4s Hartree-Fock orbitals for He($\xi_{1s}$) and V($\xi_{4s}$), with a cutoff distance at 4.0 Å. We fix $N_s = 16$ in the present potential, which yields an s-electron density at the first nearest-neighbor distance equaling to 0.01% of the corresponding d-electron density of vanadium atom.

## 2.3 Fitting technology

The potential functions were optimized by minimizing the weighted mean-squared deviation of properties from their target values by the simulated annealing method. Meanwhile, we have employed classical molecular static relaxation to obtain the stable configurations and formation energies of He-participated defects and clusters with a 1000-atom cell during the fitting processes. The objective function is expressed as following:

$$U = \sum \omega_i\, [f_i(\lambda_i) - F_i]^2, \tag{9}$$

which characterizes the goodness of fit of each individual set. In Eq. (9), the functional $f_i(\lambda_i)$ depends on the potential parameters, which should be obtained from molecular static relaxation. The target values $F_i$ are formation energies of defects obtained from *ab initio* calculations. The anneal optimization method is used to search the best values of potential parameters, where the initial temperature and the termination temperature were set as 500 and 50 K, respectively.

**2.4 Modeling parameters**

In the fitting processes, molecular statics was used to calculate the fitted values from a trial solution of Eq. 9. Bcc vanadium matrices with the size of $10a_0 \times 10a_0 \times 10a_0$ have been simulated in the paper. The most stable configurations of $He_nV$ and $He_n$ clusters were statistically obtained with 1000 random configurations. A tetrahedral interstitial and a vacancy site are taken as the center position of $He_n$ and $He_nV$ clusters, respectively. Furthermore, the additional He atoms were randomly placed within a sphere with a radius of $a_0$. Each random configuration was quenched to obtain the lowest-energy state (the most stable configuration). To ensure the reliability and accuracy of the present V-He potential, the formation energies of clusters were also calculated according to Eq. 1.

In order to predict the diffusion of He atom in vanadium crystals, the mean square displacement (MSD) and the climbing image nudged elastic band (CI-NEB) techniques were utilized [38-40]. Molecular dynamic method was performed to obtain the MSD curves in the NVT ensembles at the range from 300 to 1500 K. The step length is set as 2 femtoseconds, and the total time lasts 3 nanoseconds for each MD run. Based on the present potential, the binding energies of an interstitial He atom to small clusters were also calculated according to Eq. 2. The self-trapping behaviors in vanadium and iron were investigated with the present V-He potential and the Fe-He potential [41] to gain insight into the anomalous helium behaviors in vanadium, which were implemented with MD method at 300 K. For the self-trapping simulations, the interstitial He atoms with the concentration of 1.0% (*at.*) have randomly distributed within the blocks of $80a_0 \times 80a_0 \times 80a_0$. Twenty different models were constructed, which were firstly quenched to their most stable states and then thermally equilibrated for 30 picoseconds.

**3. Results and discussion**

3.1 Ab initio calculations of He-participated defect properties

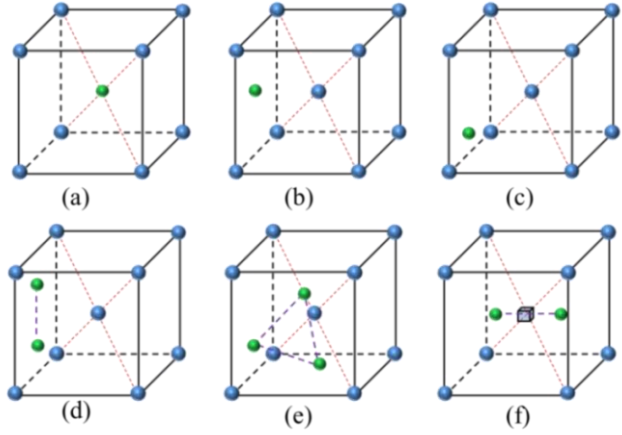

Fig. 1. (Color online) Most stable He-participated defects in bcc vanadium crystals. The blue and green balls represent the vanadium atoms and helium atoms, respectively.

Table 1. Formation energy ($E_f$) of the He-participated defects in vanadium according to *ab initio* calculations and the fitted potential description. EP represents the results with the present empirical potential.

| Configurations | Formation energy (eV) | | |
| --- | --- | --- | --- |
| | *Ab initio* results | | EP |
| | Present | Reference | |
| Sub-He | 4.63 | 4.58[25], 4.25[44] | 4.01 |
| Oct-He | 3.25 | 3.17[25], 3.24[45], 3.04[44] | 3.16 |
| Tet-He | 3.02 | 2.94[25], 3.02[45], 2.82[44] | 2.93 |
| $He_2$ | 6.10 | 5.88[43] | 5.72 |
| $He_3$ | 9.07 | 8.67[43] | 8.50 |
| $He_2$-V | 6.15 | -- | 6.14 |

In the examination of the relative stabilities of various primary $He_nV_m$ (n=1-3; m=0, 1) clusters, the main interest is always focused on the calculation of formation energies of the configurations with high symmetry. In the combination of six possible stable configurations of the He-participated defects (shown in Fig. 1), the formation energies of substitutional He atom (Sub-He), tetrahedral He atom (Tet-He), octahedral He atom (Oct-He), $He_2$, $He_3$ and $He_2$V clusters are summarized in Table 1. As one can see, the single solution He atom has the formation energies of 3.02, 3.25 and 4.63 eV at tetrahedral, octahedral and substitutional sites, respectively. He atom prefers to occupy the tetrahedral and octahedral sites. It is difficult for impurity He atom to form at the substitutional sites, which is in good agreement with the previous results[25,42-44]. As shown in Fig. 1(d), the He-He pair has the lowest formation energy of 6.1 eV when the two He atoms stay at a pair of second-nearest neighbor tetrahedral sites. The configuration of $He_3$ cluster is the most stable one only if the three He atoms locate the third-nearest neighbors tetrahedral sites (in Fig. 1(e)), with the formation energy of 9.07 eV. For the stable $He_2V$ complex (in Fig. 1(f)), it has the lowest formation energy of 6.15 eV where the primary elements distribute in the <100> direction. It has been demonstrated that He octahedral density of states (DOS) is higher than that of the tetrahedral defect at the Fermi energy, which is in agreement with the order of the preference of He in metals, namely, Tet-He, and Oct-He[25].

### 3.2 Empirical potentials

Table 2. Potential parameters of the "s-band" model for the interaction between vanadium and He atoms.

| $a_1$~$a_7$ (eV/Å$^3$) | $r_1$~$r_7$ (Å) | Other parameters | | |
|---|---|---|---|---|
| 39.880088783834 | 1.75946 | $b_1$= 0.230485525663 | eV | |
| 176.594893732905 | 1.77131 | $b_2$= 1.689739849980 | eV | |
| 3.322775563349 | 1.93780 | $b_3$= 2.619149771760 | eV | |
| -2.582702537995 | 2.18797 | $N_s$= 16.0 | 1/Å$^3$ | |
| 1.472978894497 | 2.33306 | $\xi_s$= 2.151414324324 | 1/Å | |
| -0.626688157526 | 2.77596 | $r_{ce}$= 4.0 | Å | |
| 0.171663537811 | 3.59627 | -- | | |

Based on the fitting approach outlined above, the present *ab initio* database, such as the formation energies of relaxed Sub-He, Tet-He, Oct-He, $He_2$, $He_3$ and $He_2V$ clusters (in Table 1) have been used to develop the V-He "s-band" model potential.

The key parameters of the pair potential and 's-band model' many-body interaction are given in Table 2. The pair potential, the many-body potential and the density function of the V-He interaction are presented in Fig. 2.

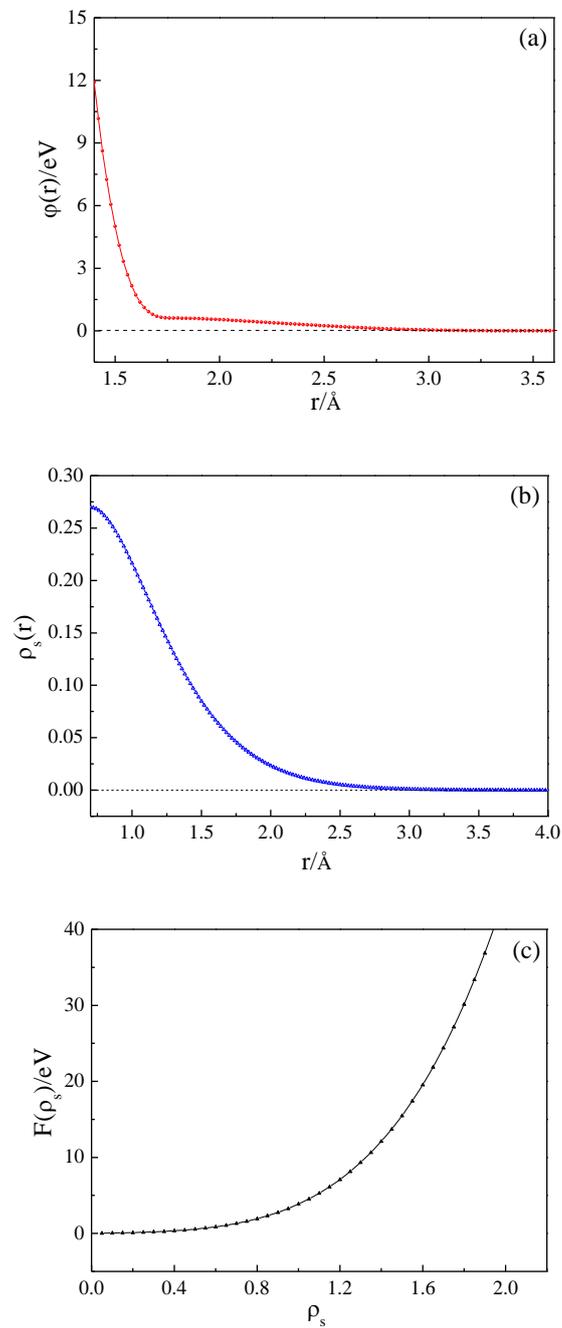

Fig. 2. (Color online) V-He interatomic potential function curves of hybridization electron density (a), the pair repulsive interaction (b), and the hybridization energy items (c).

Due to the closed-shell structure of electron of helium atoms, it is recognized by *ab initio* calculations that there is little attractive interaction between vanadium and He atoms. As a consequence, the pair potential item φ(r) should be positive within the cut-off distance (3.6 Å) shown in Fig. 2(a), which indicates the repulsive interaction. On the contrary, the previous two body V-He potential becomes negative at the distances between 2.9 and 5.3 Å [27], which indicates the existence of an unphysical attractive interaction, rather than repulsive. It is also noted that the host vanadium atoms interact little with the He atoms over the range of the second-nearest neighbor. As one can see from Fig. 2(b), the hybridization electron density $\rho_s(r)$ decreases with the increase of the distance between He and vanadium atoms. The maximum hybridization electron density (0.27) can be obtained when the distance equals to 0.7 Å. On the other hand, it can be seen from Fig. 2(c) that the hybridization energy increases with the electron density in the form of an exponent-like function. In contrast with the behaviors described in Fe-He system [41], the contribution of s-electrons to many-body interaction is very weak in V-He system, which is in agreement with *ab initio* calculations [26]. However, the hybridization energy of electrons cannot be ignored here.

To ensure the reliability and accuracy of the present V-He potential, we have checked the formation energies of the He-participated defects by molecular static method (shown in Table 1), together with the *ab initio* values for comparison. The obtained results show that the He atom prefers to occupy the tetrahedral interstitial sites with the formation energy of 2.93 eV, while it could also stably stay at the octahedral sites with the formation energy of 3.16 eV. The extra energy of 4.01 eV is needed for a He atom to substitute a host vanadium atom. It is noted that the formation energy of octahedral interstitial He atom is 0.23 eV higher than that of the tetrahedral interstitial He atom. Meanwhile, the formation energies of 5.72, 8.5 and 6.14 eV are also obtained for the $He_2$, $He_3$ and $He_2V$ clusters, respectively. The formation energies of the He-participated defects obtained with the present potential are in good agreement with the *ab initio* results [42-44]. As a consequence, the potential developed in this work provides an accurate description of a spectrum of He behaviors in vanadium.

**3.3 The anomalous behaviors of He atoms in vanadium**

It is widely known that the interstitial He atoms may migrate quickly and bind to each other, leading to the formation of helium clusters in metals. $He_n$ clusters with certain sizes may activate intrinsic defects to nucleate bubbles which could grow up by absorbing interstitial He atoms. Since the aggregation of He atoms into bubbles can induce both trans- and inter-granular brittleness even in intrinsically ductile materials [45], it is necessary to understand the dynamical processes (diffusion, clustering, and trapping) involved in bubble nucleation and growth. As the significant applications of V-He potential, a great deal of attention will be focused on the dynamical mechanism of He-participated defects in vanadium.

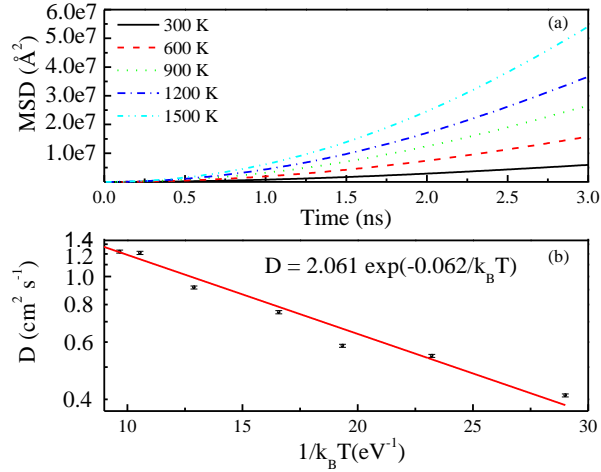

Fig. 3. (Color online) (a) The mean square displacements (MSD) as a function of time for a single interstitial helium atom at temperatures from 300 to 1500 K. (b) The diffusion coefficients of a single He atom as a function of $1/k_BT$.

In Fig. 3, we examine the capability of mean square displacement (MSD) analysis to extract reliable values of the diffusion coefficient $D$ of a single He atom undergoing Brownian motion in vacancy free vanadium. As we can see from Fig. 3(a), the MSD curves continually increase with diffusion time. The interstitial He atom has the maximum MSD values at the high temperatures (T=1500 K). The MSD curves clearly decrease with the decrease of temperature. It can be revealed that the He atom diffuse quickly among interstitial sites. Meanwhile, it is easier for a single He atom to migrate at higher temperature. Besides for the absence of pre-existing traps (vacancies, voids, grain boundaries, and so on), the only interstitial He atom is incompetent to kick-out lattice atoms to induce the generation of traps. Thus, the interstitial He atom could not be captured in our simulations.

The dependence of diffusivity on temperature is shown in Fig. 3(b). The diffusion coefficient increases with the increase of temperature, which indicates that the diffusion of He atoms will be promoted by the increase of temperature in vanadium. According to the Arrhenius equation, the migration energy barriers $E_m$ (0.062 eV) and prefactors $D_0$ (2.061 cm$^2$s$^{-1}$) can be obtained. In order to clarify the migration mechanism, we have designed two migration paths for the interstitial He atom (shown in Fig. 4): (1) from a tetrahedral site to another first-nearest neighbor tetrahedral site (T-T), and (2) from a tetrahedral site to another second-nearest neighbor tetrahedral site passing through an octahedral one (T-O-T). The climbing image nudged elastic band (CI-NEB) method is used to measure the migration energy barriers along the above paths[38]. As one can see from Fig. 4, the migration energy barriers for the T-T and T-O-T paths are 0.068 and 0.23 eV, respectively, which is in good agreement with the *ab initio* result[43]. Combining the migration energy barriers obtained from Arrhenius equation, it can be inferred that the T-T path predominates over the diffusion mechanism of interstitial He atoms in vanadium.

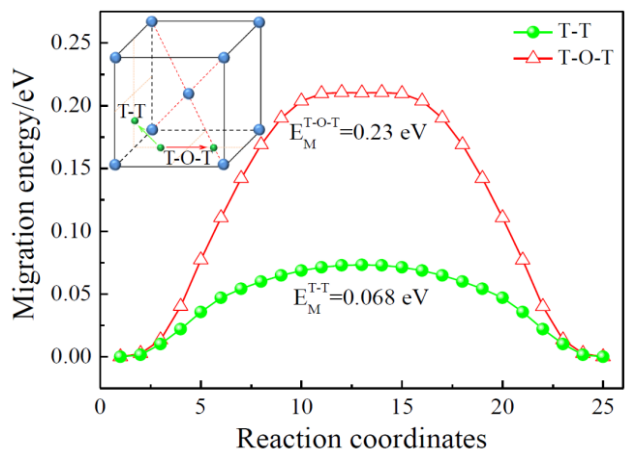

Fig. 4. (Color online) Migration of a tetrahedral interstitial He atom along the two paths T-T and T-O-T.

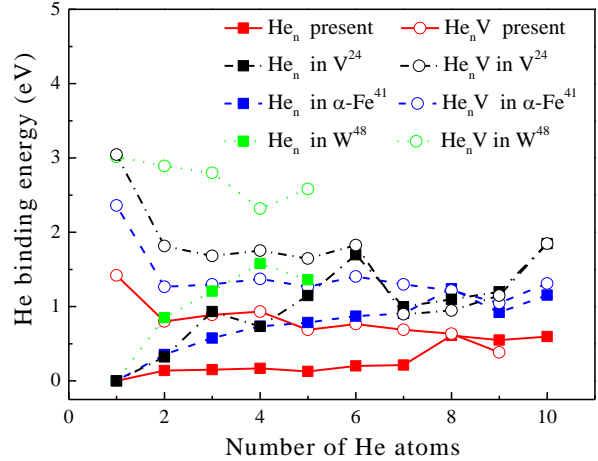

Fig. 5. (Color online) Binding energy of an additional He atom to the $He_n$ and $He_nV$ clusters as a function of the cluster size.

We have employed the present V-He interatomic potential to study the clustering behaviors of $He_nV$ and $He_n$ clusters, and their binding energies as a function of cluster size are shown in Fig. 5. Comparisons with available experimental or other calculation results in vanadium, iron, and tungsten are also presented. For bcc vanadium, the binding energies of a single He atom to the $He_n$ clusters increases with the increase of cluster size. On the contrary, the binding energies for the $He_nV$ clusters decrease when the number of He atoms increases in the $He_nV$ clusters. The previous two body V-He potential overestimates the binding energies of He atom to clusters, which is inconsistent with the experimental and *ab initio* results [42,46]. Fedorov *et al.* have performed many the thermal helium desorption spectrometry experiments in vanadium and its alloys. It was found that the $He_n$ clusters have low dissociation energies less than 0.3 eV, while the individual He atom has the migration energy of 0.13 eV. From the above mentioned results, we can infer that the single interstitial He atom binds to other He atoms or He clusters with the binding energy lower than 0.17 eV. Recently, Zhang *et al.* have also obtained the low binding energy of 0.17 eV between two helium atoms in vanadium using first principle approaches. The weak hybridization between the d state of vanadium atom and the p state of He atom is responsible for the low binding energy for He-He pairs in vanadium [26]. As one can also see from Fig. 5, the similar tendencies described in the present potential can also occur in iron and tungsten [41, 47]. However, the interstitial helium atoms tend to bind to other He atoms or helium clusters with high binding energies in Fe and W. Furthermore, the binding energies quickly increase as the number of helium atoms increases. It can be explained as that the hybridization between vanadium and He electronic states is weaker than that in Fe and W [26]. It is also noted in Fig. 5 that the same binding energy is needed to form $He_8$ and $He_8V$ clusters, which indicates that the $He_8$ cluster has transformed into the $He_8V$ cluster due to the presence of a vacancy induced by the activation of a nearby SIA.

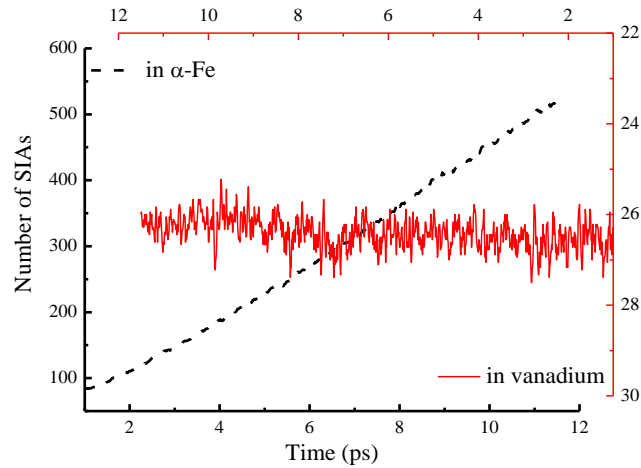

Fig. 6. (Color online) Evolution of the number of SIAs activated by the $He_n$ clusters in vanadium and α-Fe.

Based on the above mentioned binding properties of He atoms, it can be reasonably inferred that the anomalies may occur in the aggregation of He atoms in vanadium. In order to confirm the phenomenon, the self-trapping of He atoms are furtherly investigated in vanadium and iron. The number of SIAs activated by the $He_n$ clusters in Fe and vanadium are shown in Fig. 6. It is noted that a great many of SIAs (84 to 520) are quickly induced in α-Fe. In contrast with the behaviors described in iron, things become more interesting once the helium atoms dispersed in vanadium. As one can see from Fig. 6, the number of SIAs fluctuate around 27 in vanadium. It is well known that $He_n$ clusters can lead to the formation of SIAs in metals by the self-trapping mechanism in the absence of intrinsic defects. Therefore, the aggregation of He atoms could be described by the number of SIAs in vanadium and iron. It can be revealed from the obtained results that a great many of the interstitial He atoms have aggregated into clusters in iron, while only a few $He_n$ clusters were formed in vanadium. The discrepancy can be derived from the high binding energies in iron and the low binding energies in vanadium for the $He_n$ clusters (also see Fig. 5).

## 4. Conclusion

In the *ab initio* examination for the relative stabilities of various primary $He_nV_m$ (n=1-3; m=0, 1) clusters, it is found that He atom prefers to occupy the tetrahedral and octahedral sites. Due to the weak hybridization, He atoms tend to aggregate into clusters with high symmetry configurations and low binding energies. Based on the experimental and *ab initio* results, an accurate and reliable V-He interatomic potential has been developed. With the present potential, we have investigated the dynamical processes (diffusion, clustering, and trapping) relevant to the formation of bubbles.

It is found that interstitial He atoms can migrate quickly along the T-T path and they will aggregate into clusters with low binding energies in vanadium. Contrary to the $He_nV$ clusters, the binding energies of a single He atom to the $He_n$ clusters increases with the increase of cluster size. The attraction among He atoms in vanadium is weaker than that in iron and tungsten. The self-trapping behaviors of He atoms are also examined in vanadium and iron with molecular dynamics method. It is hard for He atoms to be trapped within vanadium in the absence of intrinsic defects. The trapping of He atoms depends much on the pre-existing traps, which is of great difference from that in iron.

As a final statement, it can be drawn that the present potential should be competent in the investigation for V-He system. The vanadium based materials should be resistant to the damage from the accumulation of He atoms, which is similar to from experimental examination. The present work enables us to gain insight into the essence of the V-He interaction from atomistic scales.

**Acknowledgement**

This work was supported by the "Strategic Priority Research Program" of the Chinese Academy of Sciences (Grant no. XDA03030100). Professor Fei Gao at the University of Michigan is greatly appreciated for his helpful discussions. This work was also financially supported by the National Natural Science Foundation of China (Grant Nos. 51471068, 51271075, 51371080, 11304324 and 11505266) and the Shanghai Municipal Science and Technology Commission (13ZR1448000).

**References**

[1] J.M. Chen, V.M. Chernov, R.J. Kurtz, T. Muroga, J. Nucl. Mater. 417 (2011) 289.

[2] Y. Katoh, M. Ando, A. Kohyam, J. Nucl. Mater. 323 (2003) 251.

[3] G. Kress, J. Hafner, Phys. Rev. B 48 (1993) 13115.

[4] Y. Ding, R. Ahuja, J. F. Shu, P. Chow, W. Luo, H.K. Mao, Phys. Rev. Lett. 98 (2007) 085502.

[5] Zs. Jenei, H.P. Liermann, H. Cynn, J.-H.P. Klepeis, B.J. Baer, W.J. Evans, Phys. Rev. B 83 (2011) 054101.

[6] S.J. Zinkle, Fusion Eng. Des. 74 (2005) 31.

[7] T. Muroga, J.M. Chen, V.M. Chernov, K. Fukumoto, D.T. Hoelzer, R.J. Kurtz, T. Nagasaka, B.A. Pint, M. Satou, A. Suzuki, H. Watanabe, J. Nucl. Mater. 367-370 (2007) 780.

[8] R.J. Kurtz, K. Abe, V.M. Chernov, D.T. Hoelzer, H. Matsui, T. Muroga, G.R. Odette, J. Nucl. Mater. 329-333 (2004) 47.

[9] A.F. Calder, D.J. Bacon, J. Nucl. Mater. 207 (1993) 25.

[10] D.J. Bacon, F. Gao, Y.N. Osetsky, J. Nucl. Mater. 276 (2000) 1.

[11] E. Alonso, M.J. Caturla, T.D. de la Rubia, J.M. Perlado, J. Nucl. Mater. 276 (2000) 221.

[12] F. Gao, H.L. Heinisch, R.J. Kurtz, J. Nucl. Mater. 367-370 (2007) 446.

[13] J.M. Perlado, J. Marian, D. Lodi, T.D. De La Rubia, Computer simulation of the effect of copper on defect production and damage evolution in ferritic steels[C]//MRS Proceedings. Cambridge University Press, 578 (1999) 243.

[14] N. Juslin, K. Nordlund, J. Wallenius, L. Malerba, Nucl. Instrum. Meth. B 255 (2007) 75.

[15] G. Lucas, R. Schäublin, J. Nucl. Mater. 386-388 (2009) 360.

[16] L. Malerba, D. Terentyev, P. Olsson, R. Chakarova, J. Wallenius, J. Nucl. Mater. 329-333 (2004) 1156.

[17] R. Schäublin, J. Henry, Y. Dai, CR. Phys. 9 (2008) 389.

[18] R.E. Stoller, J. Nucl. Mater. 233-237 (1996) 999.

[19] X.C. Li, Y.N. Liu, Y. Yu, G.N. Luo, X.L. Shu, G.H. Lu, J. Nucl. Mater. 451 (2014) 356.

[20] Q. Pu, Y.S. Leng, L. Tsetseris, H.S. Park, S.T. Pantelides, P.T. Cummings, J. Chem. Phys. 126 (2007) 144707.

[21] C.H. Chien, E.B. Barojas, M.R. Pederson, J. Chem. Phys. 112 (2000) 2301.

[22] S.F. Xiao, W.Y. Hu, J. Chem. Phys. 125 (2006) 014503.

[23] S.F. Xiao, W.Y. Hu, J.Y. Yang, J. Chem. Phys. 125 (2006) 184504.

[24] V.M. Chernov, V.A. Romanov, A.O. Krutskikh, J. Nucl. Mater. 271-272 (1999) 274.

[25] T. Seletskaia, Y. Osetsky, R.E. Stoller, G.M. Stocks, Phys. Rev. B 78 (2008) 134103.

[26] P.B. Zhang, T.T. Zou, J.J. Zhao, J. Nucl. Mater. 467 (2015) 465.

[27] Y. Mishin, Acta Mater. 52 (2004) 1451.

[28] S. L. Frederiksen, K.W. Jacobsen, J. Schiøtz, Acta Mater. 52 (2004) 5019.

[29] Y. Mishin, M.J. Mehl, D.A. Papaconstantopoulos, Acta Mater. 53 (2005) 4029.

[30] P.J. Hay, W.R. Wadt, J. Chem. Phys. 82 (1985) 299 .

[31] G. Kresse, J. Hafner, Phys. Rev. B 47 (1993) 558.

[32] G. Kresse, J. Furthmüller, Phys. Rev. B 54 (1996) 11169.

[33] G. Kresse, D. Joubert, Phys. Rev. B 59 (1999) 1758.

[34] J.P. Perdew, Y. Wang, Phys. Rev. B 46 (1992) 12947.

[35] M.I. Mendelev, S.Han, W.J. Son, G.J. Ackland, D.J. Srolovitz, Phys. Rev. B 76 (2007) 214105.

[36] R.A. Aziz, A.R. Janzen, M.R. Moldover, Phys. Rev. Lett. 74 (1995) 1586.

[37] E. Clementi, C. Rostti, At. Data Nucl. Data Tables 14 (1974) 177.

[38] G. Henkelman, B.P. Uberuaga, H. Jónsson, J. Chem. Phys. 113 (2000) 9901.

[39] G. Henkelman, B.P. Uberuaga, H. Jónsson, J. Chem. Phys. 113 (2000) 9978.

[40] D. Sheppard, R. Terrell, G. Henkelman, J. Chem. Phys. 128 (2008) 134106.

[41] F. Gao, H.Q. Deng, H.L. Heinisch, R.J. Kurtz, J. Nucl. Mater. 418 (2011) 115.

[42] P.B. Zhang, J.J. Zhao, Y. Qin, B. Wen, J. Nucl. Mater. 419 (2011) 1.

[43] P.B. Zhang, J.J. Zhao, Y. Qin, B. Wen, J. Nucl. Mater. 413 (2011) 90.

[44] L. J. Gui, Y.L. Liu, W.T. Wang, Y.N. Liu, K. Arshad, Y. Zhang, G.H. Lu, J.N. Yao, Chin. Mater. Res. Soc. 23 (2013) 459.

[45] M. Samaras, Mater. Today 12 (2009) 46.

[46] A.V. Fedorov, A. van Veen, A.I. Ryazanov, J. Nucl. Mater. 233-237(1996) 385.

[47] N. Juslin, B.D. Wirth, J. Nucl. Mater. 432 (2013) 61.

Highlights:

(1) An accurate and reliable interatomic potential for V-He interaction have been developed to describe the anomalous behaviors of helium atoms in vanadium.

(3) The impurity He atoms bind to other defects or clusters with low binding energies.

(4) The trapping of He atoms depends much on pre-existing defects, such as intrinsic vacancies, voids, dislocations and grain boundaries, which is of great difference from that in iron and tungsten.